\documentclass{webofc}
\usepackage[varg]{txfonts}   % Web of Conferences font
\usepackage{hyperref}
\usepackage{url}
\hypersetup{colorlinks=true,citecolor=blue,urlcolor=blue,linkcolor=blue}
\begin{document}
\begin{flushright}
DESY-24-158
\end{flushright}

\vspace{-1em}

\title{Physics case for an $e^+e^-$ collider at 500 GeV and above}

\author{\firstname{Philip} \lastname{Bechtle}\inst{1}\fnsep\thanks{\email{bechtle@physik.uni-bonn.de}} \and
        \firstname{Sven} \lastname{Heinemeyer}\inst{2}\fnsep\thanks{\email{sven.heinemeyer@cern.ch}} \and
        \firstname{Jenny} \lastname{List}\inst{3}\fnsep\thanks{\email{jenny.list@desy.de}} \and
        \firstname{Gudrid} \lastname{Moortgat-Pick}\inst{3,4}\fnsep\thanks{\email{gudrid.moortgat-pick@desy.de}} \and
        \firstname{Georg} \lastname{Weiglein}\inst{3,4}\fnsep\thanks{\email{georg.weiglein@desy.de}}
        % etc.
}

\institute{Physikalisches Institut, Rheinische Friedrich-Wilhelms-Universit\"at Bonn, Nussallee 12, Bonn, 53115, NRW, Germany
\and
Instituto de F\'isica Te\'orica (UAM/CSIC), 
Universidad Aut\'onoma de Madrid, Cantoblanco, 28049, Madrid, Spain
\and
Deutsches Elektronen-Synchrotron DESY, Notkestr.~85,  22607  Hamburg,  Germany
\and
II.\  Institut f\"ur  Theoretische  Physik, Universit\"at  Hamburg, Luruper Chaussee 149, 22761 Hamburg, Germany
          }

\abstract{Some highlights of the physics case for running an $e^+e^-$ collider at 500 GeV and above are discussed with a particular emphasis on the experimental access to the Higgs potential via di-Higgs and (at sufficiently high energy) triple Higgs production. The information obtainable from Higgs pair production at about 500 GeV is compared with the 
prospects for the HL-LHC and with the 
indirect information that can be obtained from a Higgs factory running at lower energies. 
}
\maketitle
\section{Introduction}
\label{intro}

There is wide support in the particle physics community for an $e^+e^-$ ``Higgs factory'' as a near-future new particle collider. The design options of a circular or a linear collider have important implications for the c.m.\ energy that can be reached at these facilities. While the emission of synchrotron radiation limits the currently discussed circular $e^+e^-$ colliders to energies up to about 350~GeV, significantly higher c.m.\ energies can be reached at a linear $e^+e^-$ collider. The latter design choice also offers the possibility to start with a lower-energy machine, operating for instance at a c.m.\ energy of 250~GeV, and in a second stage to perform an upgrade to higher energies, either by upgrading the accelerating structures or by extending the tunnel length. In the present article% 
\footnote{Invited plenary talk given by G.W.\ at LCWS2024, Tokyo, July 2024.}
some highlights of the physics case for running an $e^+e^-$ collider at 500 GeV and above are discussed with a particular emphasis on the direct measurement of the Higgs pair production process $e^+e^- \to Zhh$ that becomes accessible at a c.m.\ energy of at least 500~GeV (in conjunction with additional information from the weak-boson fusion process
$e^+e^- \to \nu_e\bar\nu_e hh$). 
Other important aspects of the physics programme of an $e^+e^-$ collider running at a c.m.\ energy of 500~GeV and above will only be briefly touched upon, for more details see e.g.\ the recent reports 
\cite{ILCInternationalDevelopmentTeam:2022izu,CLICdp:2018cto,Vernieri:2022fae}
and references therein.

\section{Higgs potential, Higgs self-couplings and Higgs pair production processes}

Many of the open questions of particle physics are related to the Higgs sector and in particular to the Higgs potential, which for this reason is often referred to as the ``holy grail'' of particle physics. 
In the Standard Model (SM) of particle physics a minimal form of the
Higgs potential is postulated with a single Higgs boson that is an elementary particle. 

The Higgs potential 
%is crucial for the mechanism of 
gives rise to
electroweak symmetry breaking. Thus, information about the Higgs potential plays a crucial role in determining how 
%(EWSB), 
the electroweak phase transition (EWPT) in the early universe took place. 
This in turn is important
for a possible explanation of the observed asymmetry between matter and anti-matter in the universe in terms of electroweak baryogenesis. However, the actual form of the Higgs potential that is realised in nature and its physical origin are largely unknown up to now. The Higgs potential receives contributions from the detected Higgs boson $h$ with a mass $m_h$ of about 125 GeV and from all additional scalar fields that may be present but have not been detected so far. While the bilinear term, i.e.\ the coefficient of $h^2$, is related to the measured value of $m_h$, the trilinear Higgs self-coupling, $\lambda_{hhh}$, i.e.\ the coefficient of $h^3$, and the quartic  Higgs self-coupling, $\lambda_{hhhh}$, i.e.\ the coefficient of $h^4$, are only loosely constrained so far (and only weak bounds exist on contributions to the Higgs potential from additional fields).

The existing constraints on $\lambda_{hhh}$ from the LHC have mainly been obtained from the searches for the Higgs pair production process, where in the gluon fusion channel a leading-order vertex diagram containing $\lambda_{hhh}$ and the top Yukawa coupling $y_t$ enters together with a box diagram involving the coupling factor $y_t^2$. Because of a large destructive interference between the contributions from these two diagrams the total cross section for di-Higgs production changes very substantially, by about two orders of magnitude, if $\lambda_{hhh}$ is varied around the SM value. The current upper bound on the di-Higgs production cross section from ATLAS and CMS translates into an upper limit on $\lambda_{hhh}$ that is about 7 times larger than the (tree-level) SM value (and also a lower limit can be set), assuming that all other Higgs couplings besides $\lambda_{hhh}$ are fixed to their SM values~\cite{ATLAS:2024ish,CMS:2022dwd} (this assumption can be relaxed by incorporating data from single Higgs production~\cite{ATLAS:2022jtk}). 

While the existing bounds on $\lambda_{hhh}$ are rather weak, they nevertheless already probe so far untested parameter regions of physics beyond the SM because loop corrections to 
%Higgs self-couplings 
$\lambda_{hhh}$ can be much larger, by more than two orders of magnitude, than to the couplings of $h$ to gauge bosons and fermions, see \cite{Kanemura:2002vm,Bahl:2022jnx,Bahl:2023eau} for investigations of extended Higgs sectors and \cite{Durieux:2022hbu} for SMEFT analyses. A very significant upward shift in $\lambda_{hhh}$ is also motivated in many scenarios giving rise to a strong first-order EWPT which is required for electroweak baryogenesis, see e.g.\ \cite{Biekotter:2022kgf} for the case of the 2HDM, where the parameter region featuring a strong first-order EWPT and a potentially detectable gravitational wave signal at the future space-based observatory LISA is correlated with an enhancement of $\lambda_{hhh}$ compared to the SM value by about a factor of 2.

At an $e^+e^-$ linear collider with a c.m.\ energy of at least 500~GeV the Higgs pair production process $e^+e^- \to Zhh$ (and also the weak-boson fusion process
$e^+e^- \to \nu_e\bar\nu_e hh$) can be measured directly and in a model-independent way. This is a qualitatively new feature distinguishing the physics capabilities of an $e^+e^-$ collider at 500~GeV and above from the ones of Higgs factories at lower energies such as the CEPC, the FCC-ee and a lower-energy version of an $e^+e^-$ linear collider. The impact of the direct measurement of the Higgs pair production processes on the determination of $\lambda_{hhh}$ will be discussed in the next section.

First experimental constraints on the quartic Higgs self-coupling $\lambda_{hhhh}$ can be obtained at the HL-LHC, see \cite{Stylianou:2023xit,Papaefstathiou:2023uum} for recent exploratory studies, and be further improved at future lepton colliders with c.m.\ energies beyond 1~TeV~\cite{Maltoni:2018ttu,Chiesa:2020awd,Gonzalez-Lopez:2020lpd,Stylianou:2023xit}, 
see the discussion in Sect.~\ref{sec:beyond500} below.

\section{Determination of $\lambda_{hhh}$ from the Higgs pair production processes at a 
550 GeV $e^+e^-$ collider in comparison to the prospects at the HL-LHC}

\begin{figure}[ht!]
\centering
  %\begin{center}	
  \includegraphics[width=0.8\textwidth]{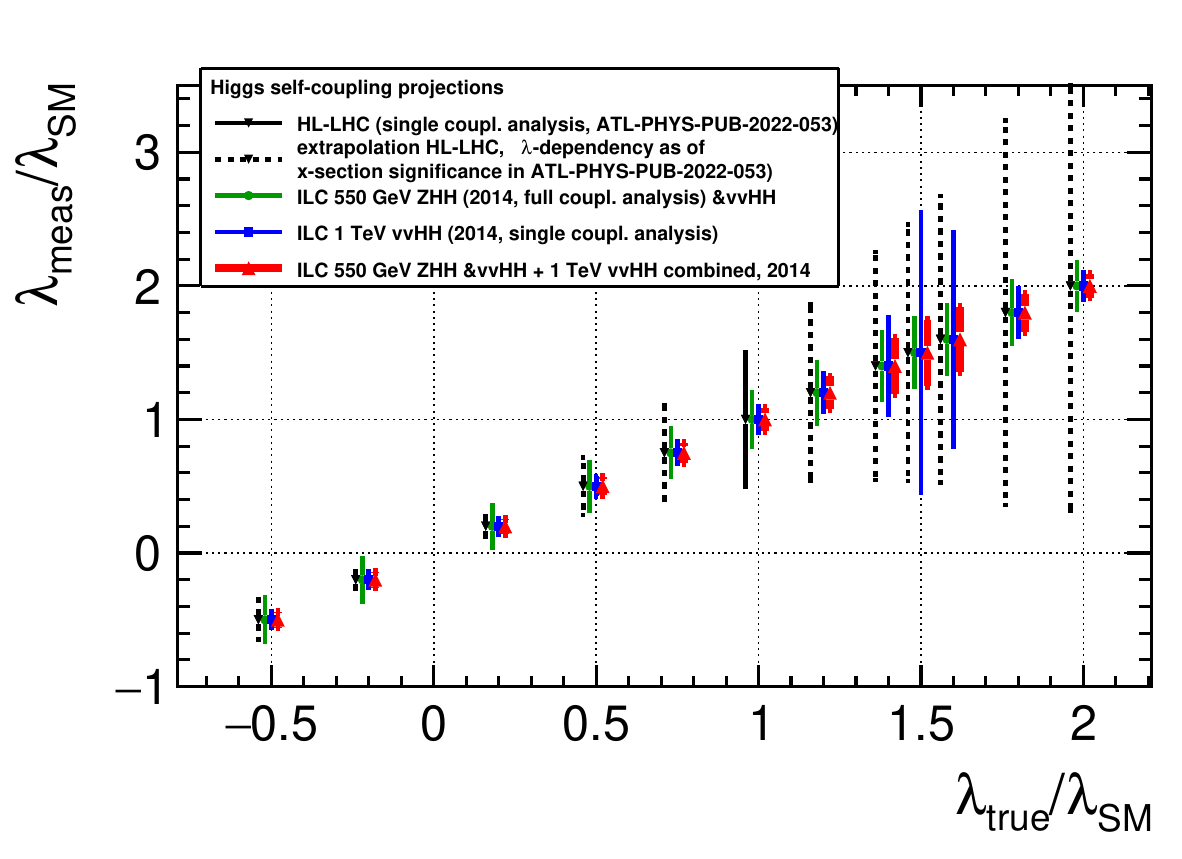}
  %\end{center}
\caption{Projected accuracies for $\lambda_{hhh}$ a the HL-LHC and a 550 GeV $e^+e^-$ collider in dependence of the actual value of $\lambda_{hhh}$ that is realised in nature~\cite{JL}.}
\label{fig:kappalambda}
\end{figure}

While the current projections for the HL-LHC (for the combined integrated luminosity that is expected to be collected by ATLAS and CMS) yield an experimental uncertainty for $\lambda_{hhh}$ of about 50\% at the 68\% C.L.\ if the SM value is realised in nature~\cite{Cepeda:2019klc}, the projected accuracy at a 550 GeV linear $e^+e^-$ collider is about 20\%, combining the results from the $Zhh$ and $\nu_e\bar\nu_e hh$ channels~\cite{JTianComb} (an update reflecting various analysis improvements is currently prepared~\cite{JLupdate}). Because of the reduced cross section as a consequence of the destructive interference of the contributing diagrams, the projected accuracy at the HL-LHC degrades to about 84\% if the value of $\lambda_{hhh}$ that is realised in nature is actually twice as large as the SM prediction (this would be favoured in scenarios giving rise to a strong first-order EWPT, see above). Since on the other hand the interference contributions in the $Zhh$ channel at a 550 GeV linear $e^+e^-$ collider are constructive, the accuracy improves to about 9\% in this case, such that the linear collider accuracy is almost an order of magnitude better than the one at the HL-LHC.%
\footnote{While it is likely that future projections for the HL-LHC will exhibit an improved prediction over the present case, due to the strong drop in the di-Higgs cross-section at the HL-LHC for values of $\lambda_{hhh}$ that are larger than the SM prediction it seems rather unlikely that the accuracy reachable at the HL-LHC in this region will be competitive with the one at an $e^+e^-$ collider with a c.m.\ energy of about 500~GeV.}
This is illustrated in Fig.~\ref{fig:kappalambda}, where for the HL-LHC a cross section extrapolation has been used. The significant degradation of the HL-LHC accuracy on $\lambda_{hhh}$ if the actual value is higher than the SM prediction has also been found in a recent projection from the ATLAS Collaboration~\cite{ATL-PHYS-PUB-2024-016}. In Fig.~\ref{fig:kappalambda} the improvements in accuracy from additional $e^+e^-$ data at 1~TeV are also displayed.

%%%%%%%%%%%%%%%%%%%%%%%%% F I G U R E %%%%%%%%%%%%%%%%%%%%%%%%%%%%%%%%%%
%%%%%%%%%%%%%%%%%%%%%%%%% F I G U R E %%%%%%%%%%%%%%%%%%%%%%%%%%%%%%%%%%

\section{Comparison with the sensitivity of $e^+e^-$ Higgs factories at lower energies}

Since the Higgs pair production processes are beyond the kinematic reach of circular $e^+e^-$ Higgs factories such as the CEPC and the FCC-ee, at those facilities one needs to resort to the indirect sensitivity of observables measured at lower energies to loop contributions involving $\lambda_{hhh}$. Specifically, the single Higgs observables receive one-loop contributions that depend on $\lambda_{hhh}$, while in the predictions for the electroweak precision observables at the Z~pole and for the W-boson mass $\lambda_{hhh}$ enters at the two-loop level. An analysis of the sensitivity to a certain parameter via loop contributions is necessarily restricted to the specific model or the specific theoretical framework that is chosen for the theoretical prediction. A well-known example is the derivation of indirect constraints on the Higgs-boson mass within the SM via a global fit of the loop contributions (the so-called ``blue band'' plot)~\cite{ALEPH:2005ab} prior to the discovery of $h$. The results of this global fit were correctly interpreted at the time as indirect constraints within a specific model (in this case the SM) rather than a ``measurement'' of the mass of a particle that at that time was not yet discovered. It should be noted that regarding the indirect determination of $\lambda_{hhh}$ via loop contributions the situation is much more difficult than for the case of the Higgs-boson mass within the SM. While the latter is a free parameter that can be freely varied within the SM while all other parameters are kept fixed, $\lambda_{hhh}$ is not a free parameter of the SM, so that a variation of $\lambda_{hhh}$ ``within'' the SM is a priori impossible. Instead, one needs to employ a consistent framework for parameterising possible deviations from the SM, such as an EFT with a complete set of higher-dimensional operators. 

In general, the loop contributions involving $\lambda_{hhh}$ compete with much larger lowest-order contributions, with other loop contributions (e.g.\ a top-quark loop) that are numerically dominant and potentially also with loop contributions arising from physics beyond the SM (BSM). The indirect sensitivity to $\lambda_{hhh}$ via loop effects is limited by the experimental errors of the considered observables as well as by the theoretical uncertainties that are induced by unknown higher-order contributions and by the experimental errors of the input parameters (in particular $\alpha_{\mathrm{em}}$, $\alpha_{\mathrm{s}}$, $m_t$, $m_b$, \ldots). 

In an EFT approach that is consistently carried out at next-to-leading order (NLO) a large number of additional EFT operators enter compared to the ones contributing at leading order. While such additional operators contributing at NLO have recently been identified~\cite{Asteriadis:2024qim,Asteriadis:2024xts}, in the EFT fits that were carried out so far to investigate the sensitivity of the CEPC and the FCC-ee to $\lambda_{hhh}$ these contributions were not yet taken into account. Furthermore, even a complete basis of dimension-6 operators at NLO in a SMEFT approach is still far from being unique in the description of possible physics scenarios. This refers on the one hand to the possible inclusion of dimension-8 (and higher-dimensional) operators and the related question of the range of validity of the EFT approach, and on the other hand to the fact that possible effects of light additional particles (below the heavy new physics scale that is used for the SMEFT expansion) are not accounted for in a SMEFT prescription. 

The discussion above implies that it is a key question to what extent the indirect determinations 
%of $\lambda_{hhh}$ 
via loop contributions at the CEPC and the FCC-ee will be able to associate a significant deviation of $\lambda_{hhh}$ that may be realised in nature correctly with a non-SM value of $\lambda_{hhh}$ in view of the plethora of other possible BSM contributions that enter at the same loop order and of the involved experimental and theoretical uncertainties. This has not been demonstrated up to now. In fact, in the EFT fits carried out in this context so far no deviations of the prospective experimental measurements from the SM predictions were considered.
%(in the analyses up to now not even statistical fluctuations of the assumed future measured values around their SM predictions were taken into account).  

While in general Higgs couplings are not directly associated with physical observables, so that a model-independent measurement of a Higgs coupling is impossible as a matter of principle at any future collider, the situation regarding the determination of $\lambda_{hhh}$ from the measurement of the Higgs pair production processes at the (HL-)LHC, an $e^+e^-$ linear collider with a c.m.\ energy of at least 500~GeV or a $\gamma\gamma$ collider with a sufficiently high c.m.\ energy~\cite{Barklow:2023ess} is much more favourable than via loop effects to lower-energy observables. This is due to the fact that $\lambda_{hhh}$ already enters at leading order and competes with much fewer contributions that enter at the same order. 
The robustness of the interpretation of the results obtained in this way within SMEFT has been demonstrated in~\cite{Barklow:2017awn}.
The importance of the direct measurement of the Higgs pair production processes at lepton colliders in this context has also been pointed out, for instance, in \cite{DiVita:2017eyz,DiVita:2017vrr,deBlas:2019rxi}.
A detailed comparison of the capabilities of $e^+e^-$ colliders operating above and below the threshold for $Zhh$ production, employing global fits for scenarios where the value of $\lambda_{hhh}$ that is actually realised in nature significantly differs from the SM prediction, is on the way~\cite{globfits}.

\section{Further guaranteed physics at 500 GeV}

Besides the direct measurements of the Higgs pair production cross sections 
$e^+e^- \to Zhh$ and $e^+e^- \to \nu_e\bar\nu_e hh$, further guaranteed measurements that can be carried out at an $e^+e^-$ linear collider with a c.m.\ energy of about 500~GeV comprise for instance the 
Higgs couplings to fermions and bosons in the single Higgs production channels $e^+e^- \to Zh$ and $e^+e^- \to \nu\bar\nu h$. These measurements profit from the increased luminosity and from the information provided by beam polarisation and by the measurements of the relevant observables at different energy stages.

Another very important process that becomes accessible is 
$e^+e^- \to t \bar t h$, for which a c.m.\ energy slightly above 500 GeV is beneficial. The direct measurement of this process is advantageous for the determination of the top Yukawa coupling in comparison to indirect constraints via loop contributions in a similar way as discussed above for the trilinear Higgs-boson self-coupling. Running an $e^+e^-$ collider at this c.m.\ energy also enables a rich programme of top-quark and electroweak physics.

\section{Guaranteed physics beyond 500 GeV}
\label{sec:beyond500}

Regarding Higgs physics, the measurements of the Higgs couplings to fermions 
and bosons profit in particular from the high statistics in the weak-boson fusion channel $e^+e^- \to \nu\bar\nu h$ at high energies. For Higgs pair production the importance of the $e^+e^- \to \nu_e\bar\nu_e hh$ channel as compared to the $e^+e^- \to Zhh$ channel also grows with increasing c.m.\ energy, but the transition to the case where the weak-boson fusion channel dominates happens at significantly higher c.m.\ energies than for the single Higgs production case (depending on the actual value of $\lambda_{hhh}$ that is realised in nature).  

A qualitatively new feature at c.m.\ energies of about 1~TeV and above is the sensitivity to the triple Higgs-boson production process 
$e^+e^- \to Zhhh$ and $e^+e^- \to \nu\bar\nu hhh$, which provides 
experimental access to the quartic Higgs-boson self-coupling $\lambda_{hhhh}$. Because of its dependence also on the trilinear Higgs-boson self-coupling $\lambda_{hhh}$ (some contributions to the triple Higgs-boson production process even involve the square of $\lambda_{hhh}$), the triple Higgs-boson production process also provides complementary information on $\lambda_{hhh}$ that can be combined with the results that are obtained from the Higgs pair production process.

%%%%%%%%%%%%%%%%%%%%%%%%% F I G U R E %%%%%%%%%%%%%%%%%%%%%%%%%%%%%%%%%%
\begin{figure}[ht!]
\centering
  %\begin{center}	
  \includegraphics[width=0.8\textwidth]{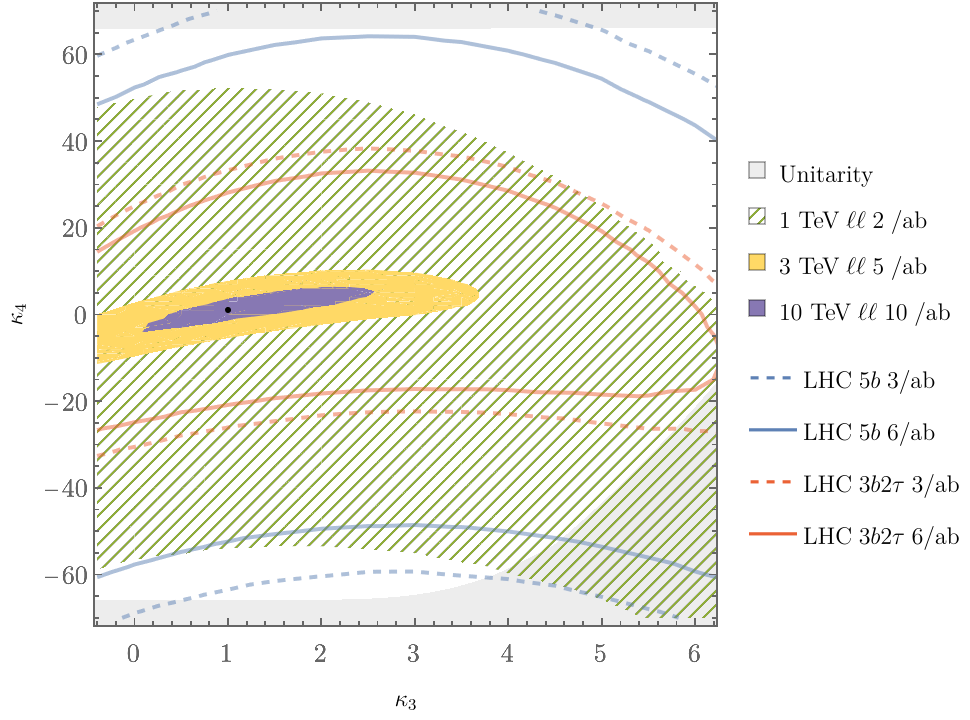}
  %\end{center}
\caption{Prospects of future lepton colliders with a c.m.\ energy of 1~TeV, 3~TeV and 10~TeV for constraining the trilinear (horizontal axis) and quartic (vertical axis) Higgs-boson self-couplings, both normalised to the tree-level values in the SM, at the 95\% C.L.\ in comparison to the projected 95\% C.L.\ contours for the $5b$ and $3b2\tau$ analyses at the HL-LHC~\cite{Stylianou:2023xit}. The shaded gray area indicates the region that is excluded by the bound from tree-level perturbative unitarity.}
\label{fig:kappa3kappa4}
\end{figure}
%%%%%%%%%%%%%%%%%%%%%%%%% F I G U R E %%%%%%%%%%%%%%%%%%%%%%%%%%%%%%%%%%

Fig.~\ref{fig:kappa3kappa4} shows the sensitivity of future lepton colliders with a c.m.\ energy of 1~TeV, 3~TeV and 10~TeV to set constraints on the trilinear (horizontal axis) and quartic (vertical axis) Higgs-boson self-couplings, both normalised to the tree-level values in the SM~\cite{Stylianou:2023xit}. The prospective sensitivities shown here are obtained just from the triple Higgs-boson production processes, i.e.\ without additional information from Higgs-boson pair production and single Higgs-boson production. The prospects for the lepton colliders are compared to the ones for the HL-LHC obtained in a recent exploratory study for the $5b$ channel and the $3b2\tau$ channel~\cite{Stylianou:2023xit}. These prospective bounds go significantly beyond the current theoretical constraints from tree-level perturbative unitarity, which are also displayed in Fig.~\ref{fig:kappa3kappa4}. The displayed results indicate that the HL-LHC is competitive to a 1~TeV lepton collider in constraining $\lambda_{hhhh}$, while the higher-energetic lepton colliders (see also \cite{Maltoni:2018ttu,Chiesa:2020awd,Gonzalez-Lopez:2020lpd}) can significantly improve on the HL-LHC capabilities.

\section{Sensitivity to new particles at 500 GeV and beyond}

An increase in the c.m.\ energy of an $e^+e^-$ collider obviously extends the kinematic reach for detecting BSM particles. A typical search process at an $e^+e^-$ collider is the pair production of new particles. It has been demonstrated by detailed studies for many examples of BSM scenarios that the obtained exclusion and discovery reach is very robust and that the properties of the detected particles can be determined with high precision, see e.g.~\cite{ILCInternationalDevelopmentTeam:2022izu}.

Concerning recent possible hints for BSM particles from the LHC searches, a hypothetical additional Higgs boson at 95~GeV~\cite{CMS:2024yhz,ATLAS:2024bjr,Biekotter:2023oen} could be copiously produced~\cite{Drechsel:2018mgd} and studied in detail at a 250 GeV $e^+e^-$ Higgs factory~\cite{Biekotter:2023jld,Heinemeyer:2021msz} if it has a sufficiently large coupling to the $Z$~boson. However, if this is not the case the most promising production channels at an $e^+e^-$ collider might be the production together with a $t \bar t$ pair or the production of a pair of the Higgs boson at 95~GeV together with a $Z$~boson (possibly via the detected Higgs boson $h$ as an intermediate state). The exploration of those channels would require a significantly higher c.m.\ energy, preferably of 500~GeV or higher. The interpretation of the very significant excess over the perturbative QCD background in the $t \bar t$ search channel observed very recently by the CMS Collaboration~\cite{CMS-PAS-HIG-22-013} in terms of a CP-odd Higgs boson at about 360~GeV could be testable (depending on the details of the production process) at an $e^+e^-$ collider with a c.m.\ energy significantly above 500~GeV (since an excess at the $t \bar t$ threshold could also be caused by BSM particles with masses far above the $t \bar t$ threshold~\cite{GWECFA}, c.m.\ energies even above 1~TeV might be required for testing such a scenario). Information from an $e^+e^-$ collider with sufficiently high c.m.\ energy could be instrumental for distinguishing the interpretation of the CMS signal in terms of a $t \bar t$ bound state from possible BSM scenarios.

\section{Conclusions}

An $e^+e^-$ collider running at a c.m.\ energy of 500~GeV or above has a very rich physics programme consisting of guaranteed measurements and a high sensitivity for detecting possible new particles. Among the guaranteed measurements the direct and model-independent measurement of the Higgs pair production process $e^+e^- \to Zhh$ is a qualitative game-changer distinguishing the physics capabilities of an $e^+e^-$ collider at 500~GeV and above from the ones of Higgs factories at lower energies such as the CEPC, the FCC-ee and a lower-energy version of an $e^+e^-$ linear collider. 

The information on the 
trilinear Higgs-boson self-coupling that can be obtained
from the Higgs pair production process is crucial for gaining experimental access to the Higgs potential, the ``holy grail'' of particle physics which is the key to many of the most pressing questions about the fabric of nature. The capabilities of an $e^+e^-$ collider with a c.m.\ energy of about 550 GeV for determining the trilinear Higgs-boson self-coupling go significantly beyond the prospective HL-LHC sensitivities. This holds for the case where the value of the trilinear Higgs-boson self-coupling that is realised in nature agrees with the SM prediction, and becomes even more pronounced for the case, typically favoured in scenarios giving rise to a strong first-order EWPT that could explain the observed asymmetry between matter and anti-matter in the universe, where the actual value of the trilinear Higgs-boson self-coupling is somewhat higher than the SM value. The determination of the trilinear Higgs-boson self-coupling from the Higgs pair production process is also superior to the indirect constraints on $\lambda_{hhh}$ that can be obtained at lower-energetic Higgs factories via loop effects involving $\lambda_{hhh}$ that compete with a large variety of other contributions entering at the same order. As furthermore discussed above, the highest-energetic lepton colliders provide sensitivity for
constraining the quartic Higgs self-coupling.

The unique capabilities in measuring the Higgs pair production processes in combination with the significantly extended reach for BSM searches are a strong motivation for designing a future $e^+e^-$ Higgs factory such that an upgrade to at least 500 GeV is possible.

% BibTeX or Biber users please use (the style is already called in the class, ensure that the "woc.bst" style is in your local directory)
\bibliography{bibliography.bib}

\begin{thebibliography}{38}

\bibitem{ILCInternationalDevelopmentTeam:2022izu}
A.~Aryshev et~al. (ILC International Development Team), {The International
  Linear Collider: Report to Snowmass 2021} (2022), \texttt{2203.07622}.

\bibitem{CLICdp:2018cto}
T.K. Charles et~al. (CLICdp, CLIC), {The Compact Linear Collider (CLIC) - 2018
  Summary Report}, \textbf{2/2018} (2018), \texttt{1812.06018}.
  \doiwoc{10.23731/CYRM-2018-002}

\bibitem{Vernieri:2022fae}
C.~Vernieri et~al., {Strategy for Understanding the Higgs Physics: The Cool
  Copper Collider}, JINST \textbf{18}, P07053 (2023), \texttt{2203.07646}.
  \doiwoc{10.1088/1748-0221/18/07/P07053}

\bibitem{ATLAS:2024ish}
G.~Aad et~al. (ATLAS), {Combination of Searches for Higgs Boson Pair Production
  in pp Collisions at s=13\,\,TeV with the ATLAS Detector}, Phys. Rev. Lett.
  \textbf{133}, 101801 (2024), \texttt{2406.09971}.
  \doiwoc{10.1103/PhysRevLett.133.101801}

\bibitem{CMS:2022dwd}
A.~Tumasyan et~al. (CMS), {A portrait of the Higgs boson by the CMS experiment
  ten years after the discovery.}, Nature \textbf{607}, 60 (2022),
  \texttt{2207.00043}. \doiwoc{10.1038/s41586-022-04892-x}

\bibitem{ATLAS:2022jtk}
G.~Aad et~al. (ATLAS), {Constraints on the Higgs boson self-coupling from
  single- and double-Higgs production with the ATLAS detector using pp
  collisions at s=13 TeV}, Phys. Lett. B \textbf{843}, 137745 (2023),
  \texttt{2211.01216}. \doiwoc{10.1016/j.physletb.2023.137745}

\bibitem{Kanemura:2002vm}
S.~Kanemura, S.~Kiyoura, Y.~Okada, E.~Senaha, C.P. Yuan, {New physics effect on
  the Higgs selfcoupling}, Phys. Lett. B \textbf{558}, 157 (2003),
  \texttt{hep-ph/0211308}. \doiwoc{10.1016/S0370-2693(03)00268-5}

\bibitem{Bahl:2022jnx}
H.~Bahl, J.~Braathen, G.~Weiglein, {New Constraints on Extended Higgs Sectors
  from the Trilinear Higgs Coupling}, Phys. Rev. Lett. \textbf{129}, 231802
  (2022), \texttt{2202.03453}. \doiwoc{10.1103/PhysRevLett.129.231802}

\bibitem{Bahl:2023eau}
H.~Bahl, J.~Braathen, M.~Gabelmann, G.~Weiglein, {anyH3: precise predictions
  for the trilinear Higgs coupling in the Standard Model and beyond}, Eur.
  Phys. J. C \textbf{83}, 1156 (2023), [Erratum: Eur.Phys.J.C 84, 498 (2024)],
  \texttt{2305.03015}. \doiwoc{10.1140/epjc/s10052-023-12173-8}

\bibitem{Durieux:2022hbu}
G.~Durieux, M.~McCullough, E.~Salvioni, {Charting the Higgs self-coupling
  boundaries}, JHEP \textbf{12}, 148 (2022), [Erratum: JHEP 02, 165 (2023)],
  \texttt{2209.00666}. \doiwoc{10.1007/JHEP12(2022)148}

\bibitem{Biekotter:2022kgf}
T.~Biek\"otter, S.~Heinemeyer, J.M. No, M.O. Olea-Romacho, G.~Weiglein, {The
  trap in the early Universe: impact on the interplay between gravitational
  waves and LHC physics in the 2HDM}, JCAP \textbf{03}, 031 (2023),
  \texttt{2208.14466}. \doiwoc{10.1088/1475-7516/2023/03/031}

\bibitem{Stylianou:2023xit}
P.~Stylianou, G.~Weiglein, {Constraints on the trilinear and quartic Higgs
  couplings from triple Higgs production at the LHC and beyond}, Eur. Phys. J.
  C \textbf{84}, 366 (2024), \texttt{2312.04646}.
  \doiwoc{10.1140/epjc/s10052-024-12722-9}

\bibitem{Papaefstathiou:2023uum}
A.~Papaefstathiou, G.~Tetlalmatzi-Xolocotzi, {Multi-Higgs boson production with
  anomalous interactions at current and future proton colliders}, JHEP
  \textbf{06}, 124 (2024), \texttt{2312.13562}.
  \doiwoc{10.1007/JHEP06(2024)124}

\bibitem{Maltoni:2018ttu}
F.~Maltoni, D.~Pagani, X.~Zhao, {Constraining the Higgs self-couplings at
  $e^+e^-$ colliders}, JHEP \textbf{07}, 087 (2018), \texttt{1802.07616}.
  \doiwoc{10.1007/JHEP07(2018)087}

\bibitem{Chiesa:2020awd}
M.~Chiesa, F.~Maltoni, L.~Mantani, B.~Mele, F.~Piccinini, X.~Zhao, {Measuring
  the quartic Higgs self-coupling at a multi-TeV muon collider}, JHEP
  \textbf{09}, 098 (2020), \texttt{2003.13628}.
  \doiwoc{10.1007/JHEP09(2020)098}

\bibitem{Gonzalez-Lopez:2020lpd}
M.~Gonzalez-Lopez, M.J. Herrero, P.~Martinez-Suarez, {Testing anomalous $H-W$
  couplings and Higgs self-couplings via double and triple Higgs production at
  $e^+e^-$ colliders}, Eur. Phys. J. C \textbf{81}, 260 (2021),
  \texttt{2011.13915}. \doiwoc{10.1140/epjc/s10052-021-09048-1}

\bibitem{Cepeda:2019klc}
M.~Cepeda et~al., {Report from Working Group 2}: {Higgs Physics at the HL-LHC
  and HE-LHC}, CERN Yellow Rep. Monogr. \textbf{7}, 221 (2019),
  \texttt{1902.00134}. \doiwoc{10.23731/CYRM-2019-007.221}

\bibitem{JTianComb}
J.~Tian, talk at the 3rd ECFA Workshop on $e^+e^-$ Higgs, Electroweak and Top
  Factories, Paris, October 2024,
  https://indico.in2p3.fr/event/32629/contributions/140462/

\bibitem{JLupdate}
J.~List et al., in preparation

\bibitem{ATL-PHYS-PUB-2024-016}
Tech. Rep. ATL-PHYS-PUB-2024-016, CERN, Geneva (2024),
  \urlstyle{tt}\url{https://cds.cern.ch/record/2910850}

\bibitem{JL}
J.~List et~al., Higgs self-coupling strategy at linear $e^+e^-$ colliders,
  ILD-PHYS-PROC-2024-018, contribution to the proceedings of ICHEP 2024,
  Prague, July 2024

\bibitem{ALEPH:2005ab}
S.~Schael et~al. (ALEPH, DELPHI, L3, OPAL, SLD, LEP Electroweak Working Group,
  SLD Electroweak Group, SLD Heavy Flavour Group), {Precision electroweak
  measurements on the $Z$ resonance}, Phys. Rept. \textbf{427}, 257 (2006),
  \texttt{hep-ex/0509008}. \doiwoc{10.1016/j.physrep.2005.12.006}

\bibitem{Asteriadis:2024qim}
K.~Asteriadis, S.~Dawson, P.P. Giardino, R.~Szafron, {Prospects for New
  Discoveries Through Precision Measurements at $e^+e^-$ Colliders} (2024),
  \texttt{2406.03557}.

\bibitem{Asteriadis:2024xts}
K.~Asteriadis, S.~Dawson, P.P. Giardino, R.~Szafron, {The $e^+ e^- \rightarrow
  Z H$ Process in the SMEFT Beyond Leading Order} (2024), \texttt{2409.11466}.

\bibitem{Barklow:2023ess}
T.~Barklow, C.~Emma, Z.~Huang, A.~Naji, E.~Nanni, A.~Schwartzman, S.~Tantawi,
  G.~White, {XCC: An X-ray FEL-based $\gamma\gamma$ Compton Collider Higgs
  Factory}, JINST \textbf{18}, P07028 (2023), \texttt{2306.10057}.
  \doiwoc{10.1088/1748-0221/18/07/P07028}

\bibitem{Barklow:2017awn}
T.~Barklow, K.~Fujii, S.~Jung, M.E. Peskin, J.~Tian, {Model-Independent
  Determination of the Triple Higgs Coupling at e+e- Colliders}, Phys. Rev. D
  \textbf{97}, 053004 (2018), \texttt{1708.09079}.
  \doiwoc{10.1103/PhysRevD.97.053004}

\bibitem{DiVita:2017eyz}
S.~Di~Vita, C.~Grojean, G.~Panico, M.~Riembau, T.~Vantalon, {A global view on
  the Higgs self-coupling}, JHEP \textbf{09}, 069 (2017), \texttt{1704.01953}.
  \doiwoc{10.1007/JHEP09(2017)069}

\bibitem{DiVita:2017vrr}
S.~Di~Vita, G.~Durieux, C.~Grojean, J.~Gu, Z.~Liu, G.~Panico, M.~Riembau,
  T.~Vantalon, {A global view on the Higgs self-coupling at lepton colliders},
  JHEP \textbf{02}, 178 (2018), \texttt{1711.03978}.
  \doiwoc{10.1007/JHEP02(2018)178}

\bibitem{deBlas:2019rxi}
J.~de~Blas et~al., {Higgs Boson Studies at Future Particle Colliders}, JHEP
  \textbf{01}, 139 (2020), \texttt{1905.03764}.
  \doiwoc{10.1007/JHEP01(2020)139}

\bibitem{globfits}
H.~Bahl et al., in preparation

\bibitem{CMS:2024yhz}
A.~Hayrapetyan et~al. (CMS), {Search for a standard model-like Higgs boson in
  the mass range between 70 and 110 GeV in the diphoton final state in
  proton-proton collisions at $\sqrt{s}$ = 13 TeV} (2024), \texttt{2405.18149}.

\bibitem{ATLAS:2024bjr}
G.~Aad et~al. (ATLAS), {Search for diphoton resonances in the 66 to 110 GeV
  mass range using $pp$ collisions at $\sqrt{s}=13$ TeV with the ATLAS
  detector} (2024), \texttt{2407.07546}.

\bibitem{Biekotter:2023oen}
T.~Biek\"otter, S.~Heinemeyer, G.~Weiglein, {95.4~GeV diphoton excess at ATLAS
  and CMS}, Phys. Rev. D \textbf{109}, 035005 (2024), \texttt{2306.03889}.
  \doiwoc{10.1103/PhysRevD.109.035005}

\bibitem{Drechsel:2018mgd}
P.~Drechsel, G.~Moortgat-Pick, G.~Weiglein, {Prospects for direct searches for
  light Higgs bosons at the ILC with 250 GeV}, Eur. Phys. J. C \textbf{80}, 922
  (2020), \texttt{1801.09662}. \doiwoc{10.1140/epjc/s10052-020-08438-1}

\bibitem{Biekotter:2023jld}
T.~Biek\"otter, S.~Heinemeyer, G.~Weiglein, {The CMS di-photon excess at 95 GeV
  in view of the LHC Run 2 results}, Phys. Lett. B \textbf{846}, 138217 (2023),
  \texttt{2303.12018}. \doiwoc{10.1016/j.physletb.2023.138217}

\bibitem{Heinemeyer:2021msz}
S.~Heinemeyer, C.~Li, F.~Lika, G.~Moortgat-Pick, S.~Paasch, {Phenomenology of a
  96~GeV Higgs boson in the 2HDM with an additional singlet}, Phys. Rev. D
  \textbf{106}, 075003 (2022), \texttt{2112.11958}.
  \doiwoc{10.1103/PhysRevD.106.075003}

\bibitem{CMS-PAS-HIG-22-013}
Tech. Rep. CMS-PAS-HIG-22-013, CERN, Geneva (2024),
  \urlstyle{tt}\url{https://cms-results.web.cern.ch/cms-results/public-results/preliminary-results/HIG-22-013}

\bibitem{GWECFA}
G.~Weiglein, talk at the 3rd ECFA Workshop on $e^+e^-$ Higgs, Electroweak and
  Top Factories, Paris, October 2024,
  https://indico.in2p3.fr/event/32629/contributions/142843/

\end{thebibliography}

\end{document}